\documentclass[aps,pra,twocolumn,amsmath,amssymb]{revtex4}
\usepackage{graphicx}
\usepackage{amsmath}
\usepackage{color}

\begin{document}

\title{Exact two-body solutions and Quantum defect theory of two dimensional dipolar quantum gas}

\author{Jianwen Jie}

\author{Ran Qi}
\email{qiran@ruc.edu.cn}

\affiliation{Department of Physics, Renmin University of China, Beijing 100872, P. R. China}
\date{\today}
\begin{abstract}
In this paper, we provide the two-body exact solutions of two dimensional (2D) Schr\"{o}dinger equation with isotropic $\pm 1/r^3$ interactions. Analytic quantum defect theory are constructed base on these solutions and are applied to investigate the scattering properties as well as two-body bound states of ultracold polar molecules confined in a quasi-2D geometry. Interestingly, we find that for the attractive case, the scattering resonance happens simultaneously in all partial waves which has not been observed in other systems. The effect of this feature on the scattering phase shift across such resonances is also illustrated.
\end{abstract}
\maketitle

\section{Introduction}
In recent years, great progress has been made in the production of ultracold polar molecules thanks to the development of stimulated Raman adiabatic passage (STIRAP) technique \cite{STIRIP1,STIRIP2}. Extensive experimental and theoretical efforts have been devoted to exploring the production of dipolar gases and to revealing exotic quantum phases resulting from the anisotropic
nature and long-range character of the dipolar interaction \cite{STIRIP3,dipolemanybody1,dipolemanybody2,dipolemanybody3,dipolemanybody4,dipolemanybody5}. In three dimensions, the system is usually suffered from large chemical reaction rate leading to a very short life time \cite{chemical reaction1,chemical reaction2,chemical reaction3}. The reaction rate, however, can be greatly suppressed by confining the system to lower dimensionality and makes it possible to realize a stable quantum many-body system \cite{chemical reaction4}. As a result, a lot of efforts have been made in order to understand both two-body and many-body properties of dipolar quantum gases in two or one dimensional geometries \cite{dipolemanybody1,2Dscattering1,2Dscattering2,2Dscattering3,2Dscattering4,2Dscattering5,2Dscattering6,2Dscattering7,2Dp+ip,1Dscattering1,1Dscattering2,1Dscattering3,1Dscattering4,1Dscattering5}. However, up to now, few analytical results are known even in the simple two-body problems due to the existence of long range $1/r^3$ tail in the interaction potential.

In this paper, following a similar method developed for the three dimensional scattering problem under repulsive $1/r^3$ and attractive $1/r^6$ interaction \cite{BoGaoSolution}, we present  exact solutions, in the form of a generalized Neumann expansion, for the 2D quantum scattering problem with either repulsive or attractive isotropic $1/r^3$ interactions. Both cases are relevant for current polar molecule experimental setups \cite{2Dscattering2,2Dp+ip}. 
With the help of these exact solutions, we will construct the analytic quantum defect theory (QDT) \cite{QDT1,QDT2,QDT3,QDT4} for realistic dipolar scattering in the quasi-2D confinement and investigate the corresponding scattering properties as well as the two-body bound states. 

We start in Sec. II by summarizing the explicit form of our exact solutions and also give the long range and short range asymptotic behavior analytically. In Sec. III, to be more self contained, we briefly review the general scattering formula in 2D and provide the definition of phase shift and cross sections. In Sec. IV, we first introduce the experimental setups under our consideration for two different kinds of dipole-dipole scattering in quasi-2D geometry. Then we construct the analytic QDT for these two cases and present our results on the scattering properties. Finally, we conclude ourselves in Sec. V.

\section{Solutions of the Schr\"{o}dinger equation}
We consider the Schr\"{o}dinger equation for $\pm 1/r^3$ type potentials in two spacial dimensions satisfied by the radial wave function $\psi_m(k,r)$:
\begin{eqnarray}
\left[-\frac{d^{2}}{d r^{2}}+\frac{m^2-1/4}{r^{2}}\pm\frac{D}{r^3}-\overline{\varepsilon}\right]u_{m}(r)=0,\label{Schrodinger}
\end{eqnarray}
where $u_{m}(r)=\psi_m(k,r)/\sqrt{r}$, $D=\mu d^2/\hbar^2$ is the dipolar length with $\mu$ and $d$ being the reduced mass and dipole moment, $\overline{\varepsilon}=2\mu \varepsilon/\hbar^{2}=k^{2}$ with $\varepsilon$ and $k$ the scattering energy and wave number, and $m\hbar$ is angular momentum. The solutions for the three dimensional version of this type of equation are already provide in the repulsive case \cite{BoGaoSolution}. We find that the method used in \cite{BoGaoSolution} can be easily generalized to two spacial dimensions as well as to the attractive case and we will present the explicit form of our exact solutions to Eq.~(\ref{Schrodinger}) in the following part of this section.

We found that there exists a pair of linearly
independent solutions with energy-independent asymptotic behaviors near the origin~($r\ll D$). The explicit form can be written as
\begin{eqnarray}
\overline{u}_{\varepsilon m}^{+1}(r) &\!=\!& \frac{G^{-1}_{\varepsilon m}(-\nu)\xi_{\varepsilon m}^{+}(r)\!-\!G^{-1}_{\varepsilon m}(\nu)\eta_{\varepsilon m}^{+}(r)}{\sin2\pi\nu},\label{up1}\\
\overline{u}_{\varepsilon m}^{+2}(r) &\!=\!&-G^{-1}_{\varepsilon m}(-\nu)\xi_{\varepsilon m}^{+}(r)\!-\!G^{-1}_{\varepsilon m}(\nu)\eta_{\varepsilon m}^{+}(r),\label{up2}\\
\overline{u}_{\varepsilon m}^{-1}(r) &\!=\!& \frac{-G^{-1}_{\varepsilon m}(-\nu)\xi_{\varepsilon m}^{-}(r)\!+\!G^{-1}_{\varepsilon m}(\nu)\eta_{\varepsilon m}^{-}(r)}{2\sin\pi\nu},\label{um1}\\
\overline{u}_{\varepsilon m}^{-2}(r) &\!=\!& \frac{G^{-1}_{\varepsilon m}(-\nu)\xi_{\varepsilon m}^{-}(r)\!+\!G^{-1}_{\varepsilon m}(\nu)\eta_{\varepsilon m}^{-}(r)}{2\cos\pi\nu}.\label{um2}
\end{eqnarray}
Here and below the plus and minus signs on the superscripts correspond to the repulsive and attractive interactions, respectively. Functions $\xi_{\varepsilon m}^{\pm}(r)$ and $\eta_{\varepsilon m}^{\pm}(r)$ in (\ref{up1})-(\ref{um2}) are another pair of linearly independent solutions that takes the form of a generalized Neumann expansion:
\begin{eqnarray}
\xi_{\varepsilon m}^{\pm}(r) &=& \sum_{n=-\infty}^{\infty}r^{1/2}b_{n}^{\pm}J_{\nu+n}(kr),\\
\eta_{\varepsilon m}^{\pm}(r)  &=& \sum_{n=-\infty}^{\infty}(-1)^{n}b_{n}^{\pm}r^{1/2}J_{-\nu-n}(kr),
\end{eqnarray}
where
\begin{eqnarray}
b_{j}^{\pm}&\!=\!&(\pm\Delta)^{j}\frac{\Gamma(\nu)\Gamma(\nu\!-\!m\!+\!1)\Gamma(\nu\!+\!m+1)}{\Gamma(\nu\!+\!j)\Gamma(\nu\!-\!m\!+\!j\!+\!1)\Gamma(\nu\!+\!m\!+\!j\!+\!1)}c_{j}(\nu),\nonumber\\
\\
b_{-j}^{\pm}&\!=\!&(\pm\Delta)^{j}\frac{\Gamma(\nu\!-\!j\!+\!1)\Gamma(\nu\!-\!m\!-\!j)\Gamma(\nu\!+\!m\!-\!j)}{\Gamma(\nu\!+\!1)\Gamma(\nu\!-\!m)\Gamma(\nu\!+\!m)}c_{j}(-\nu),\nonumber\\
\end{eqnarray}
with $j$ being a positive integer, $ \Delta=kD/2$ and
\begin{equation}
c_{j}(\nu)=Q(\nu+j-1)Q(\nu+j-2)\cdots Q(\nu)b_{0}.
\end{equation}
The coefficient $b_{0}$ is a normalization constant which can be set to 1, and $Q(\nu)$ is given by a continued fraction
\begin{equation}
Q(\nu)= \frac{1}{1-\varepsilon_{s}\frac{Q(\nu+1)}{(\nu+1)[(\nu+1)^{2}-m^2](\nu+2)[(\nu+2)^{2}-m^2]}},
\end{equation}
where $\varepsilon_{s}$ is a scaled energy defined as
\begin{equation}
\varepsilon_{s}\equiv\Delta^{2}=\frac{1}{4}\frac{\varepsilon}{(\hbar^{2}/2\mu)(1/D^{2})}.
\end{equation}
The $G_{\varepsilon m}(\nu)$ function in (\ref{up1})-(\ref{um2}) is defined as
\begin{equation}
G_{\varepsilon m}(\nu) = \Delta^{-\nu} \frac{\Gamma(1+m+\nu)\Gamma(1-m+\nu)}{\Gamma(1-\nu)}C(\nu),
\end{equation}
where $C(\nu)=\lim_{j\rightarrow \infty}c_{j}(\nu)$.

Finally, $\nu$ is a root of a characteristic function
\begin{eqnarray}
\Lambda_{m}(\nu,\varepsilon_{s})\equiv\Big(\nu^{2}-m^2\Big)-\frac{\varepsilon_{s}}{\nu}\Big[ \bar{Q}(\nu)- \bar{Q}(-\nu)\Big],
\end{eqnarray}
where $\bar{Q}(\nu)$ is defined as
\begin{equation}
  \bar{Q}(\nu)=\frac{Q(\nu)}{(\nu+1)[(\nu+1)^{2}-m^{2}]}.
\end{equation}
The solution of $\nu$ for $\Lambda_{m}(\nu,\varepsilon_{s})=0$ could either be real or complex depending on the scattering energy and angular momentum.

The pair of solutions $\overline{u}_{\varepsilon m}^{\pm1}$ and $\overline{u}_{\varepsilon m}^{\pm2}$ 
 have been defined in such a
way that they have energy-independent behavior near the origin$~$($r\ll D$), which are given as
\begin{eqnarray}
\overline{u}_{\varepsilon m}^{+1}(r)&\rightarrow&r^{3/4}\sqrt{\frac{1}{\pi \sqrt{D}}}e^{-2\sqrt{\frac{D}{r}}},\label{up1asym}\\
\overline{u}_{\varepsilon m}^{+2}(r)&\rightarrow&-r^{3/4}\sqrt{\frac{1}{\pi \sqrt{D}}}e^{2\sqrt{\frac{D}{r}}},\label{up2asym}\\
\overline{u}_{\varepsilon m}^{-1}(r)&\rightarrow&r^{3/4}\sqrt{\frac{1}{\pi \sqrt{D}}}\sin\left(2\sqrt{\frac{D}{r}}\!-\!\frac{\pi}{4}\right),\label{um1asym}\\
\overline{u}_{\varepsilon m}^{-2}(r)&\rightarrow&r^{3/4}\sqrt{\frac{1}{\pi \sqrt{D}}}\cos\left(2\sqrt{\frac{D}{r}}\!-\!\frac{\pi}{4}\right),\label{um2asym}
\end{eqnarray}
for both positive and negative energies. Note that the solution $\overline{u}_{\varepsilon m}^{+1}$ approaches zero exponentially in the limit $r\rightarrow0 $ and thus is the physical
solution for pure repulsive $1/r^3$ interaction.

For positive scattering energy $\varepsilon>0$, the asymptotic behaviors of $\overline{u}_{\varepsilon m}^{\pm1},\overline{u}_{\varepsilon m}^{\pm2}$ as  $r\rightarrow\infty$ are given as
\begin{eqnarray}
\overline{u}_{\varepsilon m}^{\pm1}(r)&\rightarrow&\frac{1}{\sqrt{2\pi k}}\left[Z_{11}^{\pm}\sin\left(kr\!-\!\frac{l\pi}{2}\right)-Z_{12}^{\pm}\cos\left(kr\!-\!\frac{l\pi}{2}\right)\right],\nonumber\\
~\label{u1asymlong}\\
\overline{u}_{\varepsilon m}^{\pm2}(r)&\rightarrow&\frac{1}{\sqrt{2\pi k}}\left[Z_{21}^{\pm}\sin\left(kr\!-\!\frac{l\pi}{2}\right)-Z_{22}^{\pm}\cos\left(kr\!-\!\frac{l\pi}{2}\right)\right],\nonumber\\
~\label{u2asymlong}
\end{eqnarray}
where $l=m-1/2$. The matrix $Z_{ij}$ are dimensionless functions of $m$ and scaled energy $\varepsilon_s$ which can be obtained analytically as
\begin{eqnarray}
Z_{11}^{+} &\!=\!& \frac{2}{\sin2\nu\pi}\Big[\frac{\alpha_{\varepsilon m}}{G_{\varepsilon m}(-\nu)}\!-\!\frac{\alpha_{\varepsilon m}\cos\pi\nu+\beta_{\varepsilon m}\sin\pi\nu}{G_{\varepsilon m}(\nu)}\Big],\nonumber\\
~\\
Z_{12} ^{+}&\!=\!& \frac{2}{\sin2\nu\pi}\Big[\frac{\beta_{\varepsilon m}}{G_{\varepsilon m}(-\nu)}\!-\!\frac{\beta_{\varepsilon m}\cos\pi\nu-\alpha_{\varepsilon m}\sin\pi\nu}{G_{\varepsilon m}(\nu)}\Big],\nonumber\\
~\\
Z_{21}^{+} &\!=\!&-2\Big[\frac{\alpha_{\varepsilon m}}{G_{\varepsilon m}(-\nu)}\!+\!\frac{\alpha_{\varepsilon m}\cos\pi\nu\!+\!\beta_{\varepsilon m}\sin\pi\nu}{G_{\varepsilon m}(\nu)}\Big]\\
~\nonumber\\
Z_{22}^{+} &\!=\!&-2\Big[\frac{\beta_{\varepsilon m}}{G_{\varepsilon m}(-\nu)}\!+\!\frac{\beta_{\varepsilon m}\cos\pi\nu\!-\!\alpha_{\varepsilon m}\sin\pi\nu}{G_{\varepsilon m}(\nu)}\Big],\\
~\nonumber\\
Z_{11}^{-} &\!=\!& \frac{1}{\sin\nu\pi}\Big[\frac{-\alpha_{\varepsilon m}}{G_{\varepsilon m}(-\nu)}\!+\!\frac{\alpha_{\varepsilon m}\cos\pi\nu\!+\!\beta_{\varepsilon m}\sin\pi\nu}{G_{\varepsilon m}(\nu)}\Big],\nonumber\\
~\\
Z_{12} ^{-}&\!=\!& \frac{1}{\sin\nu\pi}\Big[\frac{-\beta_{\varepsilon m}}{G_{\varepsilon m}(-\nu)}\!+\!\frac{\beta_{\varepsilon m}\cos\pi\nu\!-\!\alpha_{\varepsilon m}\sin\pi\nu}{G_{\varepsilon m}(\nu)}\Big],\nonumber\\
~\\
Z_{21}^{-} &\!=\!& \frac{1}{\cos\nu\pi}\Big[\frac{\alpha_{\varepsilon m}}{G_{\varepsilon m}(-\nu)}\!+\!\frac{\alpha_{\varepsilon m}\cos\pi\nu\!+\!\beta_{\varepsilon m}\sin\pi\nu}{G_{\varepsilon m}(\nu)}\Big],\nonumber\\
~\\
Z_{22}^{-} &\!=\!& \frac{1}{\cos\nu\pi}\Big[\frac{\beta_{\varepsilon m}}{G_{\varepsilon m}(-\nu)}\!+\!\frac{\beta_{\varepsilon m}\cos\pi\nu\!-\!\alpha_{\varepsilon m}\sin\pi\nu}{G_{\varepsilon m}(\nu)}\Big],\nonumber\\
\end{eqnarray}
where
\begin{eqnarray}
\alpha_{\varepsilon m} &=& \textrm{cos}[\pi(\nu-m)/2]X_{\varepsilon m}-\textrm{sin}[\pi(\nu-m)/2]Y_{\varepsilon m},~~~~\\
\beta_{\varepsilon m} &=& \textrm{sin}[\pi(\nu-m)/2]X_{\varepsilon m}+\textrm{cos}[\pi(\nu-m)/2]Y_{\varepsilon m},~~~~
\end{eqnarray}
and $X_{\varepsilon m},~Y_{\varepsilon m}$ are defined as
\begin{eqnarray}
X_{\varepsilon m} &=& \sum_{n=-\infty}^{\infty}(-1)^{n}b_{2n} ,\\
Y_{\varepsilon m} &=& \sum_{n=-\infty}^{\infty}(-1)^{n}b_{2n+1},
\end{eqnarray}
These dimensionless $Z_{ij}^+$ functions are the key quantities to calculate the scattering phase shifts which will be clear later.

For negative energy $\varepsilon<0$, $\overline{u}_{\varepsilon m}^{\pm1},\overline{u}_{\varepsilon m}^{\pm2}$ have the following asymptotic behaviors as $r\rightarrow\infty$
\begin{eqnarray}
\overline{u}_{\varepsilon m}^{\pm1}(r) &\rightarrow&\frac{1}{\sqrt{2\pi\kappa}}\Big(W_{11}^{\pm}e^{\kappa r}+W_{12}^{\pm}e^{-\kappa r}\Big),\label{u1asymneg}\\
\overline{u}_{\varepsilon m}^{\pm2}(r) &\rightarrow& \frac{1}{\sqrt{2\pi\kappa}}\Big(W_{21}^{\pm}e^{\kappa r}+W_{22}^{\pm}e^{-\kappa r}\Big),\label{u2asymneg}
\end{eqnarray}
where the $\kappa=\sqrt{-\bar{\varepsilon}}$ and $W_{ij}$ are also dimensionless functions of $m$ and $\varepsilon_s$:
\begin{eqnarray}
W_{11}^{+}&\!=\!&-\frac{D_{m}}{\sin2\pi\nu}\left[\frac{1}{i^{\nu}G_{\varepsilon m}(\nu)}\!-\!\frac{1}{i^{-\nu}G_{\varepsilon m}(-\nu)}\right],\\
W_{12}^{+}&\!=\!&-\frac{E_{m}}{2\cos\pi\nu}\left[\frac{1}{i^{\nu}G_{\varepsilon m}(\nu)}\!+\!\frac{1}{i^{-\nu}G_{\varepsilon m}(-\nu)}\right],\\
W_{21}^{+}&\!=\!&-D_{m}\left[\frac{1}{i^{\nu}G_{\varepsilon m}(\nu)}\!+\!\frac{1}{i^{-\nu}G_{\varepsilon m}(-\nu)}\right],\\
W_{22}^{+}&\!=\!&-E_{m}\left[\frac{1}{i^{\nu}G_{\varepsilon m}(\nu)}\!-\!\frac{1}{i^{-\nu}G_{\varepsilon m}(-\nu)}\right]\sin\pi\nu,\\
W_{11}^{-}&\!=\!&\frac{D_{m}}{2\sin\pi\nu}\left[\frac{1}{i^{\nu}G_{\varepsilon m}(\nu)}\!-\!\frac{1}{i^{-\nu}G_{\varepsilon m}(-\nu)}\right],\\
W_{12}^{-}&\!=\!&\frac{1}{2}E_{m}\left[\frac{1}{i^{\nu}G_{\varepsilon m}(\nu)}\!+\!\frac{1}{i^{-\nu}G_{\varepsilon m}(-\nu)}\right],\\
W_{21}^{-}&\!=\!&\frac{1}{2}\frac{D_{m}}{\cos\pi\nu}\left[\frac{1}{i^{\nu}G_{\varepsilon m}(\nu)}\!+\!\frac{1}{i^{-\nu}G_{\varepsilon m}(-\nu)}\right],\\
W_{22}^{-}&\!=\!&\frac{1}{2}E_{m}\left[\frac{1}{i^{\nu}G_{\varepsilon m}(\nu)}\!-\!\frac{1}{i^{-\nu}G_{\varepsilon m}(-\nu)}\right]\tan\pi\nu,
\end{eqnarray}
where we have defined
\begin{eqnarray}
D_{m}=\sum_{n=-\infty}^{\infty}i^n b_n,\\
E_{m}=\sum_{n=-\infty}^{\infty}(-i)^n b_n.
\end{eqnarray}

Another important function which is usually called the  $\chi$ function is defined as
\begin{equation}
  \chi^{\pm}_{m}(\varepsilon_{s})=W_{11}^{\pm}/W_{21}^{\pm}.\label{chim}
\end{equation}
This function is useful in determining binding energy of the two-body bound state which will be clear in the following sections.


Finally, from the asymptotic behavior as $r\rightarrow0$ given in (\ref{up1asym})-(\ref{um2asym}), it is easy to show that the solution pairs have the Wronskian given by
\begin{equation}
  W(\overline{u}_{\varepsilon m}^{+1},\overline{u}_{\varepsilon m}^{+2})=-\frac{2}{\pi},~~W(\overline{u}_{\varepsilon m}^{-1},\overline{u}_{\varepsilon m}^{-2})=\frac{1}{\pi},
\end{equation}
Since the Wronskian is a constant that is independent of $r$, the asymptotic forms of solutions at large $r$ should give the same result, which requires
\begin{eqnarray}
   det(Z^{+})&=&Z^{+}_{11}Z^{+}_{22}-Z^{+}_{12}Z^{+}_{21}=4 ,\\
    det(Z^{-})&=& Z^{-}_{11}Z^{-}_{22}-Z^{-}_{12}Z^{-}_{21}=2, \\
   det(W^{+})&=&W^{+}_{11}W^{+}_{22}-W^{+}_{12}W^{+}_{21}=-2 ,\\
  det(W^{-})&=& W^{-}_{11}W^{-}_{22}-W^{-}_{12}W^{-}_{21}=-1.
\end{eqnarray}
These relationships have been verified in our calculations which provides a nontrivial check for our solution.

\begin{figure}[t]
\includegraphics[bb=49bp 170bp 547bp 619bp, height=2.9 in, width=3.2 in]
{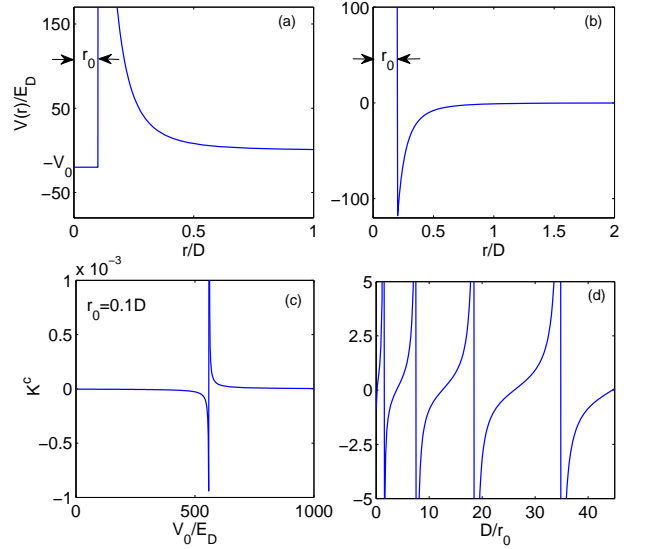}
\caption{Potential curve (upper panel) and corresponding quantum defect $K^c$ (lower panel) for two model potentials $V_+(r)$ (left collum) and $V_-(r)$ (right collum). $E_D=\hbar^2/(mD^2)$ is the characteristic energy scale associate with the dipole length $D$.}\label{Kc}
\end{figure}

\section{general scattering theory of two-body problem in two dimension}
In this section, to make our following discussions more self contained, we will briefly review the general theory of purely 2D elastic scattering under arbitrary interaction potential that decays faster than $1/r^2$. At inter-particle distance $r\rightarrow\infty$ the wave function of two distinguishable colliding atoms is represented as a superposition of the incident plane wave and scattered circular wave
\begin{eqnarray}
\psi_{\textbf{k}}(\textbf{r}) &\simeq& e^{i\textbf{k}\cdot\textbf{r}}+f(k,\theta)\frac{e^{ikr}}{\sqrt{r}},\label{scattering}
\end{eqnarray}
where $\textbf{k}$ is the relative momentum of the two particles under scattering, $f(k,\theta)$ is the scattering amplitude and $\theta$ is the angle between $\textbf{r}$ and $\textbf{k}$.

When expanding in the partial wave channels, we have
\begin{eqnarray}
  \psi_{\textbf{k}}(\textbf{r}) &=&\sum_{m=-\infty}^{\infty}\psi_{m}(k,r)e^{im\theta}\\
  f(k,\theta) &=& \sum_{m=-\infty}^{\infty}f_{m}(k)e^{im\theta}
\end{eqnarray}
where $f_{m}(k)$ and $\psi_{m}(k,r)$ are $m-$wave scattering amplitude and radial wave function, and we have
\begin{eqnarray}
\psi_{m}(k,r) &=&\psi_{m}^{0}(k,r)+f_{m}(k)\frac{e^{ikr}}{\sqrt{r}}
\end{eqnarray}
where $\psi_{m}^{0}(k,r)=i^{m}J_{m}(kr)$ is $m$ partial wave component of incident plane wave $e^{i\textbf{k}\cdot\textbf{r}}$.
Finally, using the asymptotic form of Bessel function $J_{m}(x)$ in the $x\rightarrow\infty$ limit:
\begin{eqnarray}
J_{m}(x) &\rightarrow& \sqrt{\frac{2}{\pi x}}\sin\left(x-\frac{l\pi}{2}\right)
\end{eqnarray}
where $l=m-1/2$, and thus we have
\begin{eqnarray}
 \psi_{m}(k,r) &\rightarrow&i^{m}\sqrt{\frac{2}{\pi kr}}\left[\sin\left(kr-\frac{l\pi}{2}\right)-\frac{f_{m}(k)}{4}e^{i(kr-l\pi/2)}\right]\nonumber\\
 &\propto&\sin\left[kr-\frac{l\pi}{2}+\delta_{m}(k)\right]\label{defdelta}
\end{eqnarray}
where $\delta_{m}(k)$ is the $m$ partial wave scattering phase shift which is related to scattering amplitude as
\begin{eqnarray}
f_{m} (k)&=& \sqrt{\frac{2i}{\pi k}}\frac{1}{\cot\delta_{m}(k)-i}
\end{eqnarray}
The 2D total and partial cross section for two distinguishable particles are thus given as
\begin{eqnarray}
\sigma(k)&=&\sigma_{0}(k)+2\sum_{m=1}^{\infty} \sigma_{m}(k)\label{sigmatot}\\
\sigma_{m}(k)&=& 2\pi|f_{m}(k)|^{2}=\frac{4}{k}\sin^2\delta_{m}(k)\label{sigmam}
\end{eqnarray}
In the case of identical particles, the scattering wave function in Eq.~(\ref{scattering}) should be symmetrized(anti-symmetrized) for bosons(fermions) and the cross sections have an extra factor 2 in the r.h.s. of Eq.~(\ref{sigmam}). In this case, only even(odd) partial wave has nonzero contributions for bosonic(fermionic) particles.

\begin{figure}[t]
\includegraphics[bb=29bp 280bp 563bp 517bp, height=1.6 in, width=3.4 in]
{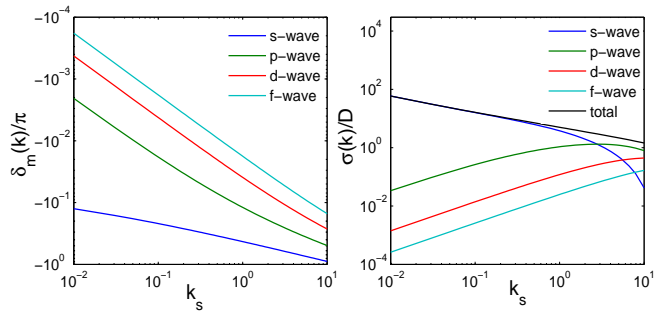}
\caption{Scattering phase shifts (left pannel) and cross sections (right pannel) of pure repulsive $1/r^3$ interaction for the first four partial waves.}\label{Pure_repulsive}
\end{figure}

\section{quantum defect theory for quasi-2D dipole-dipole scattering}
We consider two different cases of dipole-dipole scattering when two polar molecules are strongly confined along $z-$axis with a confining length $a_{\bot}\ll D$, while
moving freely in the $x-y$ plane. Case I : by applying a strong static electric field perpendicular to the $x-y$ plane, all dipole moments are aligned along $z-$axis.
In this case the interaction between two polar molecules has an isotropic repulsive $D/r^3$ tail \cite{2Dscattering1}. Case II : one implements a strong electric field fast rotating
 within the $x-y$ plane and thus generates a fast rotating dipolar moment in each polar molecule. In this case, the dipole-dipole interaction has an
 isotropic attractive $-D/r^3$ tail \cite{2Dp+ip}. In both cases, the inter molecular interaction will only deviate from the simple $\pm D/r^3$ form at short distance when
 $r\lesssim a_{\bot}\ll D$ \cite{2Dscattering1}. As a result, the effect of this deviation can be encoded in a simple short range boundary condition
\cite{QDT1,QDT2,QDT3,QDT4}. This justifies the implementation of quantum defect theory which will be constructed below.

\begin{figure*}
  \centering
  \includegraphics[bb=14bp 169bp 566bp 606bp, scale=0.8]{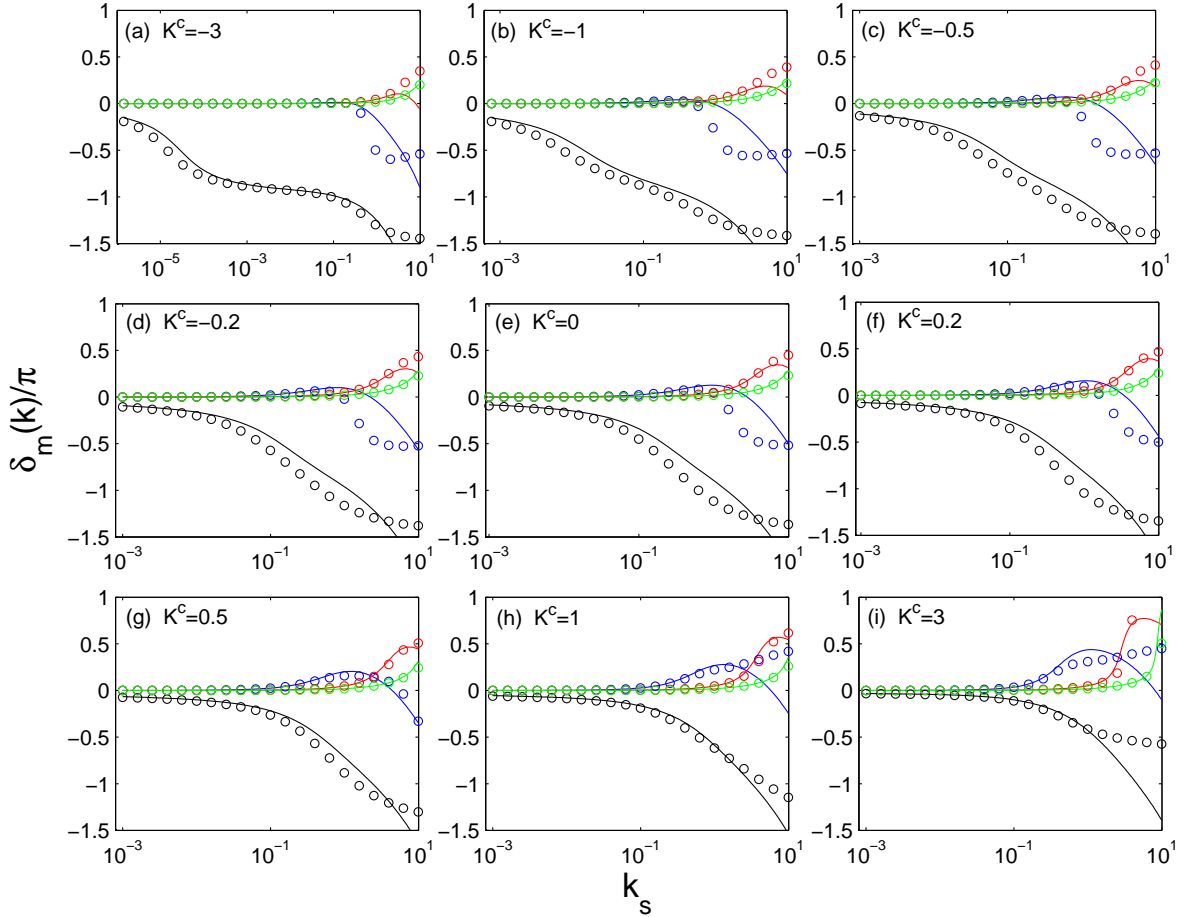}\\
  \caption{Phase shifts of first four partial waves for $-D/r^{3}$ interaction across the scattering resonance. Solid lines refer to exact results from Eq. (\ref{deltaattr}) while open circles are from low energy expansion in Eq. (\ref{delta0attr})-(\ref{delta3attr}).}\label{attractivephase}
\end{figure*}

For any two dimensional interaction with a long range $\pm D/r^3$ tail, the radial part of scattering wave function in $m$-partial wave channel with energy $\varepsilon$ can be generally written as $\psi_{m}(k,r)=u_{\varepsilon m}(r)/\sqrt{r}$ with $u_{\varepsilon m}(r)$ given as
\begin{eqnarray}
u_{\varepsilon m}(r) &=& A_{\varepsilon m}[\overline{u}_{\varepsilon m}^{1}(r)-K^{c}\overline{u}_{\varepsilon m}^{2}(r)]
\end{eqnarray}
where $K^c$ is usually called the quantum defect. At positive scattering energy, combining the asymptotic behavior of $\overline{u}_{\varepsilon m}^{1,2}(r)$ given in (\ref{u1asymlong}), (\ref{u2asymlong}) with the definition of phase shift in (\ref{defdelta}), one immediately obtains the phase shift as
\begin{equation}
K_{m}\equiv\tan\delta_{m}=\frac{K^{c}Z_{22}-Z_{12}}{Z_{11}-K^{c}Z_{21}}.\label{generaldelta}
\end{equation}

As for two-body bound states, one should use the asymptotic behavior at negative energy as given in (\ref{u1asymneg}) and (\ref{u2asymneg}). Since a physical bound state must decay exponentially at large $r$, the binding energy $E_b$ can be determined by requiring the coefficient the of $e^{\kappa r}$ term in $u_{\varepsilon m}(r)$ to vanish. This leads to the following equation for $E_b$
\begin{equation}
\chi_{m}(\varepsilon_{s})=K^{c}
\end{equation}
where the $\chi_{m}(\varepsilon_{s})$ has been defined earlier in Eq. (\ref{chim}).

As analyzed above, since the interacting potential only deviate from $\pm D/r^3$ within some short distance $r_0\ll D$, $K_c$ will be determined by the boundary condition at $r_0$ which is insensitive to neither energy nor angular momentum.
In Fig. \ref{Kc} we first illustrate how $K^c$ changes with the short range behavior of interacting potentials.
We consider two different model potentials $V_{\pm}(r)$ with $\pm D/r^3$ tails but truncated at $r_0$ as shown in Fig. \ref{Kc}(a) and (b).
For $V_+(r)=-V_0\theta(r_0-r)+\theta(r-r_0)D/r^3$ we fix $r_0$ at $0.1D$ and change the short range potential depth $V_0$, while for $V_-(r)=+\infty\theta(r_0-r)-\theta(r-r_0)D/r^3$
we just change the truncation radius $r_0$ at which a hard wall boundary condition is implemented.

From Fig. \ref{Kc}, one can see that $K^c$ has very different behaviors as one tunes
the short range behaviors of the potential with repulsive and attractive $1/r^3$ tail. In the repulsive case, as shown in Fig. \ref{Kc}(c),
$K^c$ is nearly zero everywhere except in the vicinity of some extremely narrow shape resonances.
This behavior is mainly due to the existence of a large repulsive barrier which makes the wave function almost unaffected by the short range attractive part of the potential.
 In contrast, for potential with an attractive tail the value of $K_c$ changes significantly and experiences a sequence of much wider resonances as one tunes the short range behavior, see \ref{Kc}(d). As a result, for the repulsive case, we will only consider the scattering in pure repulsive limit corresponding to $K^c=0$, while for the attractive case, we investigate both scattering and bound state properties across a shape resonance where $K^c$ can be tuned from $-\infty$ to $\infty$.

\subsection{Scattering for pure repulsive $1/r^3$ potential}
In this case, one can set $K^c=0$ and the phase shift is given as
\begin{equation}
\tan\delta^+_{m}=-\frac{Z^+_{21}}{Z^+_{11}}.
\end{equation}
In Fig. \ref{Pure_repulsive}, we show the results of phase shift and partial cross section for the first four partial waves. One can see clearly that, in the low energy regime $kD\ll1$ the s-wave channel completely dominates over higher partial waves, while at higher energy when $kD\gtrsim1$ all partial waves has non-negligible contributions. At very low scattering energy, the asymptotic behavior of phase shift can be obtained analytically from our exact solution. Below we provide the low energy expansion of $\tan\delta^+_{m}$ for the first a few partial waves:
\begin{eqnarray}
\tan\delta^+_0&\simeq&\frac{\pi}{2}\frac{1+(4/\pi)k_s\ln(k_s\bar{a}^+_0)}{\ln(k_s\bar{a}^+_0)-\pi k_s}\label{delta0rep}\\
\tan\delta^+_1&\simeq&-\frac{2k_s}{3}\left[1+\frac{3\pi}{16}k_s\ln(k_s\bar{a}^+_1)-\frac{4}{5}k_s^2\right]\\
\tan\delta^+_2&\simeq&-\frac{2k_s}{15}\left(1-\frac{\pi}{32}k_s+\frac{52}{1575}k_s^2\right)\\
\tan\delta^+_3&\simeq&-\frac{2k_s}{35}\left(1-\frac{\pi}{128}k_s+\frac{236}{99225}k_s^2\right)\label{delta3rep}
\end{eqnarray}
where $k_s=kD$, $\bar{a}^+_0=\exp(3\gamma)/2$ is simply the dimensionless 2D s-wave scattering length and $\bar{a}^+_1=\exp(3\gamma-11/12)/2$ with $\gamma$ being the Euler's constant.
For all $m>0$ partial waves, the leading order behavior agrees with that from the first order Born approximation:
\begin{equation}
\tan\delta^+_{m}\simeq-\frac{2k_s}{4m^2-1},
\end{equation}
while the sub leading terms in Eq. (\ref{delta0rep})-(\ref{delta3rep}) are new in this work.
These analytic expressions provide very good estimation for the phase shifts at low energy $k_s\lesssim1$ as shown in Fig. \ref{Pure_repulsive} (dashed lines)
and will be useful in many-body calculations \cite{2Dp+ip}.

\subsection{Scattering and bound state for interaction with an attractive $-1/r^3$ tail}
In this case, a quantum defect $K^c$ is required to fix the scattering property as well as the bound states. The $m-$partial wave scattering phase shift is given by Eq. (\ref{generaldelta}) as
\begin{equation}
\tan\delta^-_{m}=\frac{K^{c}Z^-_{22}-Z^-_{12}}{Z^-_{11}-K^{c}Z^-_{21}},\label{deltaattr}
\end{equation}
In Fig. \ref{attractivephase}, we show the scattering phase shift for the first four partial waves at different quantum defect $K^c$.

\begin{figure}[t]
\includegraphics[bb=28bp 170bp 551bp 609bp, height=3.0 in, width=3.4 in]
{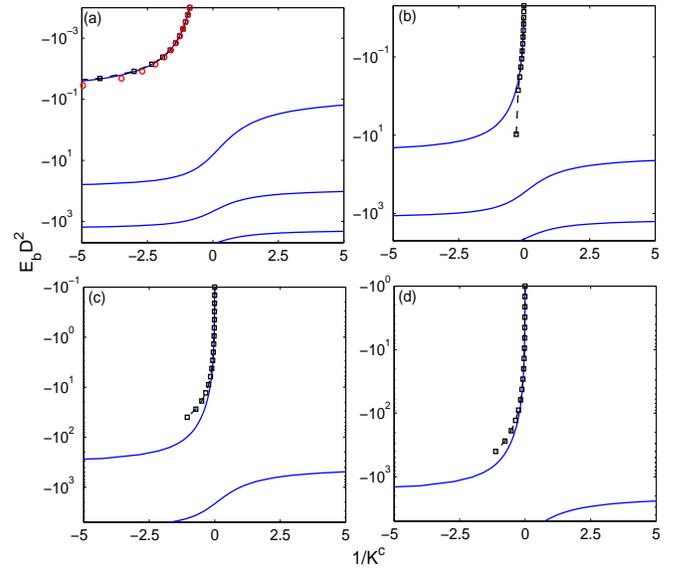}
\caption{Binding energy for the first four partial waves where (a)-(d) refers to $m=0-4$. Solid line, open squares and open circles refer to numerical solution of (\ref{Ebattractive}), analytic formulas (\ref{Eb0})-(\ref{Eb3}), and the leading s-wave behavior $E_b=-1/a_0^2$.} \label{Attractive_Bound_state}
\end{figure}

Below we provide the low energy expansion of $\tan\delta^-_{m}$ for the first a few partial waves:
\begin{eqnarray}
\tan\delta^-_0\!&\simeq&\!\frac{\pi}{2}\frac{1-\frac{4}{\pi}k_s\ln|k_s\bar{a}^-_0|}{\ln|k_s\bar{a}^-_0|+\pi k_s}\label{delta0attr}\\
\tan\delta^-_1&\simeq&
\frac{2k_s}{3}\left[1-\frac{\frac{3\pi}{16}\ln|k_s\bar{a}^-_1|\!+\!\frac{4}{5}k_s}{1-\Big(\frac{\ln^2|k_s\bar{a}^-_1|}{8}-B^c\Big)k_s^2}k_s \right]\label{delta1attr}\\
\tan\delta^-_2&\simeq&
\frac{2k_s}{15}\left[1+\frac{\frac{\pi}{32}\!+\!\frac{52}{1575}k_s}{1+\frac{\ln|k_s\bar{a}^-_2|}{120}k_s^2}k_s \right]\\
\tan\delta^-_3&\simeq&
\frac{2k_s}{35}\left[1+\frac{\frac{\pi}{128}\!+\!\frac{236}{99225}k_s}{1+\frac{\ln|k_s\bar{a}^-_3|}{1120}k_s^2}k_s \right]\label{delta3attr}
\end{eqnarray}
where $\bar{a}^-_m=\exp(3\gamma-\pi K^c-\beta_m)/2$ with $\beta_0=0$, $\beta_1=11/12$, $\beta_2=23/6$ and $\beta_3=243/40$. Comparing with the threshold behavior of phase shift for an s-wave contact interaction, one can see that the s-wave scattering length is still well defined through Eq. (\ref{delta0attr}) and is given as $a_0=\bar{a}^-_0D$. The constant $B^c$ is related to $K^c$ as
\begin{equation}
B^c=\frac{199}{1152} +\frac{5+12(K^{c})^2}{96}\pi^2.
\end{equation}
Again, the leading order behavior agrees with that from the first order Born approximation for all $m>0$ partial waves:
\begin{equation}
\tan\delta^-_{m}\simeq\frac{2k_s}{4m^2-1}.
\end{equation}
The sub leading $k_s^2\ln|k_s\bar{a}^-_1|$ term in Eq. (\ref{delta1attr}) also agrees with an earlier result obtained from a perturbative approach in \cite{2Dp+ip}.

As for two-body bound states, the binding energy for $m$ partial wave is determined by
\begin{equation}
\chi^-_{m}(\varepsilon_{s})=K^{c}.\label{Ebattractive}
\end{equation}
The low energy expansion for $\chi_m$ leads to the following approximate equation for the near threshold binding energy $E_b=-\kappa^2$ in the limit $-K^c\gg1$:

For s-wave, the binding wave number $\kappa$ satisfies
\begin{eqnarray}
&&\frac{\Omega_{\kappa}}{\pi}\!-\!\big[2\Omega_{\kappa}(\Omega^2_{\kappa}\!+\!\pi^2)\!+\!15\!-\!18\zeta(3)\big]\frac{(1\!+\!2\kappa_s)\kappa_s^2}{4\pi}\!\simeq\!K^c\nonumber\\
~\label{chi0}\label{Eb0}
\end{eqnarray}
where $\kappa_s=\kappa D$, $\Omega_{\kappa}=\ln(\kappa_s e^{3\gamma}/2)$, and $\zeta(s)$ is the Riemann Zeta function. The leading order behavior in the limit $a_0\rightarrow\infty$ is simply $E_b=-1/a_0^2$. This is consistent with the fact that the s-wave scattering length is well defined through low energy behavior of $m=0$ phase shift.

For higher partial waves, we find:
\begin{eqnarray}
E_b^{(1)}&\simeq&\frac{8}{\pi K^c t}\left(1+\frac{\ln t}{t-1} \right)^{-1}\\
E_b^{(2)}&\simeq&\frac{120}{\pi K^c}\\
E_b^{(3)}&\simeq&\frac{1120}{\pi K^c}\label{Eb3}
\end{eqnarray}
where $t=\ln|\pi K^c/2|-6\gamma+11/6$ and $E_b^{(m)}$ refers to binding energy for $m$ partial wave. In Fig. \ref{Attractive_Bound_state}, we show the results
for the first a few bound states from numerically solving (\ref{Ebattractive}) and compare with the analytic formulas in Eq. (\ref{Eb0})-(\ref{Eb3}).
From these analytic behaviors for near shreshold binding energy,
one can clearly see that the scattering resonances happen at $K^c\rightarrow\infty$ for all partial waves, at which a zero energy bound state appears for each partial wave channel. This resonant feature is also reflected in the scattering phase shifts as a rapid phase change at low scattering energy near the resonance, as illustrated in Fig. 3. It takes place at large negative and large positive value of $K^c$ for s-wave and higher partial waves, respectively. The location of this phase change corresponds to the pole of Eq. (\ref{delta0attr})-(\ref{delta3attr}), which for s-wave simply gives to $k\sim 1/a_0$. Such features in two-body scattering phase shift suggests that near the scattering resonance, the non-zero partial waves may still have important contribution to the many-body interaction energy even in the dilute regime when $nD^2\ll1$ where $n$ is the particle density. Such influences in many-body physics will be left for future studies.


\section{conclusion}
In this work, we for the first time present analytical solutions for 2D Schr\"{o}dinger equation with both repulsive and attractive inverse cubic interactions. We constructed quantum defect theory base on these solution, and investigated the scattering properties and two-body bound states of two polar molecules confined in quasi-2D geometry with two different experimental setups. We provide both exact numerical and simple analytic low energy formula for the phase shifts and two-body binding energies, which could be useful in future many-body studies. In the attractive case, we identified a resonant feature that took place simultaneously in all partial wave channels which could have important effects in corresponding many-body system.

{\it Acknowledgements.} We thank Peng Zhang, Paul S. Julienne and Hui Zhai for useful discussions. This work is supported by the Fundamental Research Funds for the Central Universities, and the Research Funds of Renmin University of China under Grant No. 15XNLF18.

\end{document}